\documentclass[aps,prb,reprint,superscriptaddress]{revtex4-2}
\usepackage{cancel}
\usepackage{graphicx}
\usepackage{bm}
\usepackage{bbold}
\usepackage[breaklinks=true, colorlinks, citecolor=blue, urlcolor=blue]{hyperref}
\usepackage{tikz}
\usepackage[compat=1.1.0]{tikz-feynman}
\usepackage{amsmath}
\usepackage{csquotes}
\usepackage[english]{babel}
\usepackage{braket}
\usepackage[dvipsnames]{xcolor}

\AtBeginDocument{%
  \renewcommand{\language}[1]{}%
}

\begin{document}

\newcommand{\black}{\color{black}}
\newcommand{\blue}{\color{BlueViolet}}
\newcommand{\red}{\color{RedViolet}}

\newcommand{\dd}{\mathrm{d}}
\newcommand{\be}{\begin{eqnarray}}
\newcommand{\ee}{\end{eqnarray}}
\newcommand{\beq}{\begin{equation}}
\newcommand{\eeq}{\end{equation}}
\newcommand{\nn}{\nonumber\\}
\newcommand{\pr}{^{\prime}}
\newcommand{\ph}{^{\text{ph}}}
\newcommand{\ct}{^{\text{CT}}}
\newcommand{\vf}{v_{\text{F}}}
\newcommand{\FD}{\mathcal{F}}
\newcommand{\vare}{\varepsilon}
\newcommand{\nf}{n_{\text{F}}}
\newcommand{\fe}{f_{e}}
\newcommand{\fh}{f_{h}}
\newcommand{\kb}{k_{\text{B}}}
\newcommand{\unit}{\sigma_{0}}
\newcommand{\tr}{\operatorname{Tr}}
\renewcommand{\Re}{\operatorname{\mathfrak{Re}}}
\renewcommand{\Im}{\operatorname{\mathfrak{Im}}}

\preprint{APS/123-QED}


\title{Anomalous thermoelectric Hall response of interacting 2D Dirac fermions}

\author{A. Daria Dumitriu-I.}
\email{alexandra-daria.dumitriu-iovanescu@outlook.com}
\affiliation{Department of Physics and Astronomy, University of Manchester, Manchester M13 9PL, United Kingdom}
\author{Feng Liu}%
\affiliation{Department of Physics, The Hong Kong University of Science and Technology, Clear Water Bay, Hong Kong, China}%
\author{Alexander E. Kazantsev}%
\affiliation{Department of Physics and Astronomy, University of Manchester, Manchester M13 9PL, United Kingdom}%
\author{Alessandro Principi}%
\email{alessandro.principi@manchester.ac.uk}
\affiliation{Department of Physics and Astronomy, University of Manchester, Manchester M13 9PL, United Kingdom}%

\date{\today}

\begin{abstract}

We study the anomalous thermoelectric Hall response of two-dimensional massive Dirac fermions to first order in the electron-electron interaction. We compute both the Nernst response to a Luttinger-type gravitational potential and the particle magnetization, the latter being required to remove spurious non-transport contributions. We show that, for arbitrary interactions, the magnetization is described by a remarkably simple formula. Surprisingly, and contrary to expectations, subtracting the magnetization currents does not make the thermoelectric Hall coefficient vanish in the zero-temperature limit.
We attribute this to violation of locality on the smallest length scales, which is inevitable in a quantized field theory, that happens to manifest itself in infrared physics. 
\end{abstract}


\maketitle

\section{Introduction\label{sec:intro}}


Thermoelectric Hall transport is a key probe of the interplay between charge and heat currents in systems that break time-reversal symmetry.  In topologically non-trivial bands with finite Berry curvature, the off-diagonal thermoelectric coefficient $L^{12}_{xy}$ (which contributes to the anomalous Nernst effect \cite{Ziman1960}) captures particle flow perpendicularly to an applied temperature gradient (or, equivalently, to an applied free fall acceleration within the Luttinger formalism \cite{luttinger1964}). When the Fermi liquid description is valid, $L^{12}_{xy}$ can be related to the electrical Hall conductivity $L^{11}_{xy}$ via the Mott relations~\cite{Ziman1960, xiao2006berryphase}. Violation of these often points to existence of special points in the energy dispersion with enhanced Berry curvature and/or density of states \cite{sakai2018, minami2020, nakamura2021, minami2024, Sakai2020}, which makes $L^{12}_{xy}$ a valuable tool in the study of band structure. Moreover, the thermoelectric coefficient can help detect non-vanishing Berry curvature when measurement of anomalous Hall conductivity only shows a small signal~\cite{nokydec2018}. 

In recent years, $L^{12}_{xy}$ has therefore become central to discussions of anomalous Nernst and thermal Hall effects in Dirac and Weyl materials \cite{liang2013, liang2017, liu2017, reichlova2018, liang2018, papaj2021, liu2017, chen2022, xu2022, Pan2022, wang2023}, as well as topological magnets \cite{Lee2004,miyasato2007, ajaya2016, xiaokang2017, guo2017, ikhlas2017, sakai2018, Xu2019, Sakai2020, mende2021, guan2026, Nakatsuji2015, nokyjun2018, nakamura2021, minami2020, li2023, tang2024, minami2024, li2025, yan2026, gong2025} serving as a benchmark for theoretical descriptions that couple thermodynamics, topology, and many-body physics.

Since time reversal in the system is broken, computing $L^{12}_{xy}$ using the standard Kubo formula comes with a pitfall that the current–current correlator carries spurious circulating (non-transport) contributions. These need to be removed \cite{smrcka1977transport, cooper1997thermoelectric} or else the result will imply that a temperature gradient induces an infinite response at vanishing temperature, which is clearly unphysical. A procedure proposed in \cite{qianniu2011energy} eliminates these circulating-current contributions and yields the true transport coefficient $(L^{12}_{xy})_{\text{tr}}$, which vanishes at $T=0$ as expected. This method, however, relies on locality. Hence, even though it gives correct results for a non-interacting system, it is not clear if it can be carried over to higher orders in perturbation theory once interactions are turned on. 

The reason is twofold. On the one hand, no physical interaction amongst electrons is truly local. On the other hand, using a fictitious delta-like contact interaction results in corrections to transport coefficients that are usually divergent and need to be regularized. This means that some level of non-locality has to be introduced on short (ultraviolet) length scales. Thus, a thorough investigation of higher order corrections, which is currently absent in the literature, is required. 

In this work, we perform such a study on a continuum model of interacting two-dimensional massive Dirac fermions. Our core result is that, even with the contact interaction $V(\bm{r}-\bm{r}')=V_c\delta(\bm{r}-\bm{r}')$, modeling overscreened Coulomb repulsion, naively satisfying the requirements of Ref.~\cite{qianniu2011energy}, the magnetisation correction no longer removes the $T \to 0$ divergence, {\it i.e.,} the supposedly true coefficient  $(L^{12}_{xy})_{\text{tr}}$ does not vanish at zero temperature.

The structure of this paper is as follows: In Section \ref{sec:description_of_the_model} we introduce the model and core definitions. In Section \ref{sec:non-interacting} we give the result for particle current magnetization $M_{N}$ and $(L_{xy}^{12})_{\mbox{tr}}$ when there are no interactions present. In Section \ref{sec:first_order} we provide the result for $M_{N}$ and $(L^{12}_{xy})_{\text{tr}}$ to first order in the interactions and point out the remaining source of divergence at $T=0$. In Section \ref{sec:discussion} we discuss the results and possible ways of resolving this difficulty and in Section \ref{sec:conclusion} we draw some conclusions. Some of the details of the calculations are placed in the Appendix.

\section{Description of the model\label{sec:description_of_the_model}}

\noindent\begin{figure}[t]
    \includegraphics[width=\linewidth]{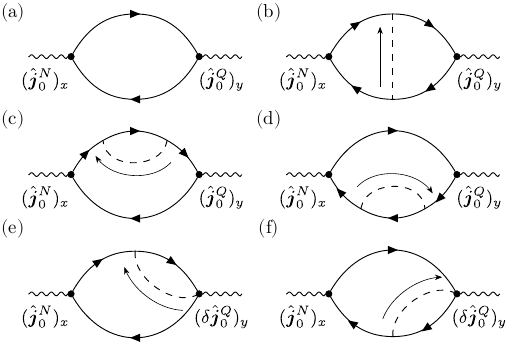}
    \caption{Diagrams contributing to $K^{12}_{xy}$ defined in Eq.~(\ref{eq:kubo}) to first order in the interaction. Panel (a) shows the non-interacting bubble diagram, consisting of two single-particle propagators (solid lines) attached to a heat-current vertex $(\hat{\bm{j}}^Q_0)_y$ (right) and a particle-current vertex $(\hat{\bm{j}}^N_0)_x$ (left). 
    Panel~(b) shows the exchange diagram, in which the density–density interaction line (dashed) connects the two propagators across the bubble.
    Panels (c) and (d) illustrate the self-energy corrections, where one of the propagators in the bubble is dressed by a single Fock insertion.  Panels (e) and (f) display the heat current vertex corrections, arising from the interaction-dependent part of the heat-current operator. In all diagrams, solid lines represent the non-interacting Matsubara Green’s functions $G^{(0)}(\bm k,i\omega_n)$, dashed lines denote the instantaneous density–density interaction $V_{\bm q}$, and dots indicate the insertion of current vertices.}
    \label{fig:1}
\end{figure}

We study a continuum model of interacting two-dimensional massive Dirac fermions, restricted to a single flavour (no real spin) and confined to the $xy$-plane. In momentum space, the system is described by the Hamiltonian
\begin{align}\label{eq:hamiltonian}
\hat{\mathcal{H}}={}&\sum_{\alpha\beta,\bm k}\hat{\psi}^{\dagger}_{\alpha,\bm k}(\bm d(\bm k)\cdot \bm \sigma_{\alpha\beta}-\mu \delta_{\alpha\beta}) \hat{\psi}_{\beta,\bm k}\nn
&+\frac{1}{2}\sum_{\bm q\neq0}\sum_{\alpha\bm k,\beta\bm k\pr}\hat{\psi}^{\dagger}_{\alpha,\bm k-\frac{\bm q}{2}}\hat{\psi}^{\dagger}_{\beta,\bm k\pr+\frac{\bm q}{2}}V_{\bm q}\hat{\psi}_{\beta,\bm k\pr-\frac{\bm q}{2}}\hat{\psi}_{\alpha,\bm k +\frac{\bm q}{2}},
\end{align}
where the first term represents the non-interacting Dirac Hamiltonian \(H_0(\bm k)=\bm d(\bm k)\cdot\bm \sigma\), while the second term describes the electron–electron interaction. The operators $\psi_{\alpha,\bm k}$ and $\psi^{\dagger}_{\alpha,\bm k}$ annihilate and create, respectively, a fermion with momentum $\bm k$ and pseudospin  $\alpha$. In the non-interacting single-particle Hamiltonian $H_0$, \(\bm d(\bm k)=(\vf k_x, \vf k_y, \Delta)\), \(\bm \sigma=(\sigma_x, \sigma_y,\sigma_z)\), $\vf$ is the Fermi velocity, $\Delta$ is half the band gap, and $\hbar=1$ is set throughout. The conduction and valence bands have, respectively, energies $\pm \vare_{\bm{k}}$, with  $\varepsilon_{\bm{k}}=\sqrt{(\vf \bm{k})^2+\Delta^{2}}$. The non-interacting Matsubara Green's function (MGF) has the form $G^{(0)}(\bm k, i\omega_{n})=\left[(i\omega_{n}+\mu)\unit-H_0(\bm k)\right]^{-1}$, where $i\omega_{n}$ is a fermionic Matsubara frequency and $\sigma_{0}$ is the $2\times2$ identity matrix, which will be later used to construct both the non-interacting and first-order interacting response functions. Lastly, we keep the interaction non-local with explicit momentum dependence in $V_{\bm{q}}$, such that this is analytic in the vicinity of $\bm{q}=0$ and falls off at infinity fast enough. This is necessary to keep all the integrals finite. At the same time, we make sure that the current density operators obey the the scaling laws of Ref.~\cite{qianniu2011energy} in the limit when $V_{\bm{q}}$ is just a constant, which we take in the end.   

Particle current response to a temperature gradient equals 
\begin{equation}\label{eq:response}
 (\bm{j}_N)_i=(L^{12}_{ij})_{\text{tr}}\bigg(-\dfrac{\partial_j T}{T}\bigg), 
\end{equation}
or equivalently
\begin{equation}\label{eq:alpha}
    (\bm{j}_N)_i=\alpha_{ij}(-\partial_j T),
\end{equation}
with $\alpha_{ij}=(L_{ij}^{12})_{\text{tr}}/T$ and $(L_{ij}^{12})_{\text{tr}}$  given by (see Ref. \cite{qianniu2011energy})
\begin{equation}\label{eq:master_formula}
(L_{ij}^{12})_{\text{tr}}=K^{12}_{ij}+\varepsilon_{ijk}(\bm{M}_N)_k.
\end{equation}
Here $K_{ij}^{12}$ is equal to
\begin{equation}
\label{eq:kubo}
K^{12}_{ij}=\lim_{\eta\to0^+}\beta\int_0^\infty dt\; e^{-\eta t}\lim_{\bm{q}\to 0}\Big\langle (\hat{\bm j}^N_{\bm{q}})_i(t); (\hat{\bm j}^Q_{-\bm{q}})_j(0)\Big\rangle V^{-1},
\end{equation}
where the Kubo canonical correlation function $\langle \hat{A};\hat{B}\rangle$ for any two observables $\hat{A}$ and $\hat{B}$ is defined as \cite{kubo_statistical_phys_2}
\begin{align}
\Big\langle\hat{A};\hat{B}\Big\rangle={}&\int_0^1 dx \mbox{Tr}\Big(\rho_0^{1-x}\hat{A}\rho_0^x \hat{B}\Big)\nn{}={}&\beta^{-1}\int_0^\beta d\lambda \mbox{Tr}\Big[\rho_0 \hat{A}(-i\lambda)\hat{B}(0)\Big],
\end{align}
with $\rho_0$ the equilibrium density matrix. Operators $\hat{j}_{\bm{q}}^N$ and $\hat{\bm{j}}_{-\bm{q}}^Q$ are the particle and heat currents\footnote{We define the Fourier transform as $f_{\bm{q}}=\int d\bm{r} f(\bm{r})e^{-i\bm{q}\cdot\bm{r}}$}, the latter defined as $\hat{\bm{j}}^Q=\hat{\bm{j}}^E-\mu \hat{\bm{j}}^N$, where $\hat{\bm{j}}_E$ is the energy current and $\mu$ is the chemical potential. Also above $V$ is the sample volume  and taking the limit $\bm q \to 0$ before $\eta \to 0^+$ makes sure that the system does not have enough time to reach equilibrium via diffusion \cite{luttinger1964}. 

Instead of directly evaluating Eq. \eqref{eq:kubo}, it is more practical to use the formula
\begin{equation}\label{eq:handy_formula}
K_{ij}^{12}=-\lim_{\omega\to0}\dfrac{(\Pi^R)^{12}_{ij}(\omega)-(\Pi^R)^{12}_{ij}(0)}{i\omega},
\end{equation}
where 
\begin{equation}\label{eq:real_bubble}
    (\Pi^R)^{12}_{ij}=-i\int_0^\infty dt\, e^{-i\omega t-\eta t}\Big\langle\Big[\big(\hat{\bm{j}}^N_{0}\big)_i(t), \big(\bm{\hat{j}}^Q_{0}\big)_j(0)\Big]\Big\rangle V^{-1},
\end{equation}
because this can be represented as a sum of Feynman diagrams, see \cite{LL}. Note that the limit $\bm{q}\to 0$ is taken here first. 
The diagrammatic representation of the non-interacting contribution to $K_{xy}^{12}$ is shown in Fig.~\ref{fig:1} panel (a), while panels (b–f) depict the first-order interaction corrections, including the exchange-like interaction (b), Fock self-energy insertions (c) and (d), and current vertex corrections associated with the interaction part of the heat current (e) and (f).

The second term in Eq. \eqref{eq:master_formula}, the St\v{r}eda term (see Refs. \cite{smrcka1977transport, cooper1997thermoelectric,  qianniu2011energy}), accounts for circulating equilibrium currents through the particle magnetisation $\bm{M}^N$, which is related to the static density–current response as
\begin{equation}
    \label{eq:streda}
    -\frac{\partial \bm M^N}{\partial \mu}=\frac{\beta}{2i}\lim_{\bm q\to0}\bm \nabla_{\bm q}\times \left\langle \hat{n}_{-\bm q}; \hat{\bm j}^N_{\bm q}\right\rangle V^{-1},
\end{equation}
where $\hat{n}_{-\bm{q}}$ is the particle density operator. Thus, the measurable thermoelectric Hall coefficient is the magnetisation-corrected combination $(L^{12}_{xy})_{\mathrm{tr}}=K^{12}_{xy}+M^N_z$, which guarantees that the response stays finite at $T=0$. Next, we need to identify the microscopic current operators entering these correlators.

The particle (number) and energy currents are each defined from their corresponding continuity equations, \(\partial_t \hat{n}+\bm \nabla \cdot\hat{\bm j}^{N}=0\) and \(\partial_t \hat{h}+\bm \nabla \cdot \hat{\bm j}^{E}=0\), where $\hat{n}$ and $\hat{h}$ are the particle and energy densities, respectively, and $\partial_t \hat{O}=i[\hat{\mathcal{H}}, \hat{O}]$. This does not fix the current densities uniquely, however. Indeed, to any $4$-vector $\hat{j}^\mu$ satisfying the continuity equation $\partial_\mu \hat{j}^\mu=0$, one can add a term $\partial_\nu \hat{M}^{\mu\nu}$, with $\hat{M}^{\mu\nu}=-\hat{M}^{\nu\mu}$ being an antisymmetric tensor. This will neither break the continuity equation nor change the value of the charge $\int d\bm{r} \hat{j}^0(\bm{r})$. 

A way to fix the densities and the corresponding currents so that the true response coefficient $(L_{ij}^{12})_{\text{tr}}$ is given by Eqs.~\eqref{eq:master_formula} and \eqref{eq:streda} is to make sure that under the following change of Hamiltonian density
\begin{equation}
\hat{h}(\bm{r}) \to \hat{h}^{\varphi,\psi}(\bm{r})=(1+\psi(\bm{r}))(\hat{h}(\bm{r})+\varphi(\bm{r})\hat{n}(\bm{r}))
\end{equation}
the currents satisfying the continuity equations also change as 
\begin{align}\label{eq:scaling_law_particle}
\hat{\bm{j}}_N(\bm{r}) &\to \hat{\bm{j}}^{\varphi,\psi}_N(\bm{r})=\big(1+\psi(\bm{r})\big)\hat{\bm{j}}_N(\bm{r}),\\ 
\hat{\bm{j}}_E(\bm{r})&\to \hat{\bm{j}}_E^{\varphi,\psi}(\bm{r})=\big(1+\psi(\bm{r})\big)^2\big(\hat{\bm{j}}_E(\bm{r})+\varphi(\bm{r})\hat{\bm{j}}_N(\bm{r})\big)\label{eq:scaling_law_energy}\nn
\end{align}
for any two functions $\varphi$ and $\psi$\footnote{Note that if this condition is not satisfied this does not mean that the true transport coefficient does not exist, it just means that it may be given by something else rather than Eqs. \eqref{eq:master_formula} and \eqref{eq:streda}. Of course on a lattice this cannot be satisfied at all, so a different procedure is needed, see \cite{kapustin2020thermal}}. Note that to check if this requirement is fulfilled one only needs to check if the following commutation relations are satisfied
\begin{align}\label{eq:commutators_h_n}
i[\hat{h}(\bm{r}), \hat{n}(\bm{r}')]&=-\bm{\nabla}_{\bm{r}'}\cdot\Big[\hat{\bm{j}}_N(\bm{r})\delta(\bm{r}-\bm{r}')\Big],\\
\big[\hat{n}(\bm{r}), \hat{n}(\bm{r}')\big]&=0
\end{align}
and
\begin{equation}\label{eq:commutators_h_h}
i[\hat{h}(\bm{r}),\hat{h}(\bm{r}')]=-\Big(\bm{\nabla}_{\bm{r}'}-\bm{\nabla}_{\bm{r}}\Big)\cdot\Big[\hat{\bm{j}}_{E}(\bm{r})\delta(\bm{r}-\bm{r}')\Big],
\end{equation}
which can be seen by direct substitution of these into the continuity equations. 

Note that because we will not study heat transport in this work, we do not actually need Eqs. \eqref{eq:scaling_law_energy} and \eqref{eq:commutators_h_h} to be satisfied for us to use Eqs. \eqref{eq:master_formula} and \eqref{eq:streda}.  With a nonlocal interaction in the energy density, this is impossible. Therefore, we will only make sure that Eqs.~\eqref{eq:scaling_law_particle} and \eqref{eq:commutators_h_n} are valid, which can be done for any nonlocal interaction.   

Since the total Hamiltonian entering the continuity equation includes the interaction term, the energy-current operator $\hat{\bm j}^E$ contains both a kinetic contribution, present already in the non-interacting limit, and an interaction-dependent correction.  Physically, the kinetic part describes the usual flow of single-particle energy states, whereas the interaction-dependent part accounts for the transfer of energy between interacting quasiparticles.  The latter ensures that the total energy is conserved when the interactions are included and gives rise to diagrams (e) and (f) in Fig. \ref{fig:1}.  These play an essential role in the proper renormalisation of the response functions discussed in the following sections.

The particle density, energy density and particle current density operators that are perfectly compatible with Eqs. \eqref{eq:scaling_law_particle} and \eqref{eq:commutators_h_n} are defined in first quantised form as 
\begin{align}
    \hspace{-0.1cm}\hat{n}_{\bm{q}}={}&\sum_{i}e^{-i\bm{q}\cdot\hat{\bm{r}}_i}\\
    \hspace{-0.1cm}\hat{h}_{\bm{q}}={}&\sum_{i}\Big(e^{-i\bm{q}\cdot\hat{\bm{r}}_i}\hat{H}^0_i(\hat{\bm{p}}_i)+\hat{H}^0_i(\hat{\bm{p}_i})e^{-i\bm{q}\cdot\bm{r}_i}\Big)\nn
    &+\dfrac{1}{2}V^{-1}\sum_{\bm{k}}\hat{n}_{\bm{q}-\bm{k}}\hat{n}_{\bm{k}}V_{\bm{k}}\\\label{eq:particle_current}
    \hspace{-0.1cm}\hat{\bm{j}}^N_{\bm{q}}={}&\dfrac{1}{2}\sum_{i}\Big(\hat{\bm{v}}_ie^{-i\bm{q}\cdot\hat{\bm{r}}_i}+e^{-i\bm{q}\cdot\hat{\bm{r}_i}}\hat{\bm{v}}_i\Big)=\sum_{i}\hat{\bm{v}}_ie^{-i\bm{q}\cdot\hat{\bm{r}}_i},
\end{align}
where index $i$ labels the particles, $\hat{\bm{r}}_i$ and $\hat{\bm{p}}_i$ denote, respectively, the position and momentum operators of the $i$th particle, $\hat{H}^0_i=\vf \hat{\bm{\sigma}}_i\cdot\hat{\bm{p}}_i$ is its kinetic energy, and $\hat{\bm{v}}_i=\vf\hat{\sigma}_i$ its velocity. With the energy density defined in this way, the energy current at $\bm{q}=0$, which is all that is needed here, turns out to be
\begin{align}\label{eq:energy_current}
    \hat{\bm{j}}_{0}^E={}&\dfrac{1}{2}\sum_i\Big(\hat{\bm{v}}_i\hat{H}^0_i(\hat{\bm{p}}_i)+\hat{H}_i^0(\hat{\bm{p}}_i)\hat{\bm{v}}_i\Big)\nn{}
    &+V^{-1}\sum_{\bm{k}}\Big(\hat{n}_{-\bm{k}}\hat{\bm{j}}^N_{\bm{k}}V(\bm{k})+\dfrac{1}{2}\big(\bm{k}\cdot\hat{\bm{j}}^N_{\bm{k}})\hat{n}_{-\bm{k}}\bm{\nabla}_{\bm{k}}V(\bm{k})\Big).\nn
\end{align}

Note that the interaction should be sufficiently local to allow the definition of a local Hamiltonian density ({\it i.e.,} the range of the interaction should be smaller than the typical length-scale on which externally applied fields change significantly). We will keep the range of the interaction finite to prevent Feynman diagrams from blowing up, subtract the divergences and take the limit of a contact interaction. This naturally arises in electron liquids when the Thomas--Fermi or Debye inverse screening length is larger than a typical momentum transfer between two Dirac quasiparticles. Then, the interaction effectively reduces to a contact-like density–density term. In our previous work \cite{dumitriu-i.2024firstorder}, we showed that this interaction yields corrections that are surprisingly similar to those produced by an unscreened Coulomb interaction for certain effective fine-structure constants, which further motivates its use here.

\section{Non-interacting calculation\label{sec:non-interacting}}

The non-interacting limit plays an important role in setting the stage for the interacting calculation. It provides both a clean demonstration of the Kubo–St\v{r}eda formalism and a benchmark against which the impact of interactions can be assessed. In this case, the evaluation of $K_{xy}^{12}$ is straightforward, since only the bare bubble diagram given in Fig. \ref{fig:1}(a) contributes, with the propagators remaining undressed, which gives 
\begin{align}
\label{eq:K_12_0}
(K^{12}_{xy})^{(0)}
&=\mu \Delta \vf^2 \int_{\bm{k}}\frac{1-f(\vare_{\bm{k}}-\mu)-f(\vare_{\bm{k}}+\mu)}{2\vare_{\bm{k}}^{3}},
\end{align}
where $\int_{\bm{k}}=\int d\bm{k}/(2\pi)^2$ and $f(\varepsilon_{\bm{k}}\mp\mu)=1/(1+e^{(\varepsilon_{\bm{k}}\mp\mu)/T})$ is the electron/hole Fermi--Dirac distribution function. Here and throughout this section, the superscript $(0)$ indicates the zeroth order in the interaction.

The Kubo response originates entirely from the Berry curvature $\pm\vf^2\Delta/2\varepsilon_{\bm{k}}^3$ of the valence/conduction band, as expected for an anomalous Hall-type response. However, this expression immediately reveals a pathology: as $T$ approaches zero, $(K_{xy}^{12})^{(0)}$ generally approaches a finite, non-vanishing value, which would subsequently lead to a divergence in Eq. \eqref{eq:response}, if the second term in Eq. \eqref{eq:master_formula} were absent.

This divergence is clearly unphysical and reflects the fact that the Kubo formula, when applied directly to energy or heat currents, captures not only transport currents but also equilibrium circulating currents, which do not contribute to net transport. To isolate the genuine transport response, one must therefore subtract the appropriate magnetisation contribution, which is obtained from the static density–current response using the St\v{r}eda formula Eq. \eqref{eq:streda}. 

\begin{figure}[t]
    \includegraphics[width=\linewidth]{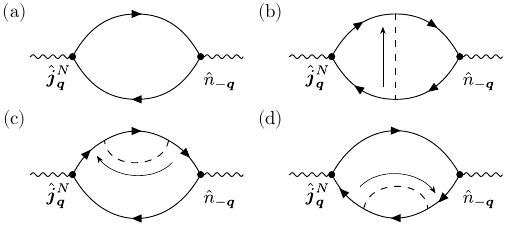}
    \caption{
    Diagrams contributing to the right-hand side of Eq. \eqref{eq:streda} to first order in the interactions. Panel (a) shows the zeroth order bubble. Panel (b) shows the exchange diagram. Panels (c) and (d) depict diagrams with self-energy insertions. The external frequency vanishes and the external momentum is kept finite.
    }
    \label{fig:2}
\end{figure}

Diagrams contributing to the static density-current response to first order in the interaction are depicted in Fig. \ref{fig:2}. Note that after they are evaluated, Eq. \eqref{eq:streda} has to be integrated with respect to $\mu$ to give $M_z^N$. As the initial value surface for $M_z^N$, we take the line $\mu=0$. Due to particle-hole symmetry, implemented by the transformation $\psi\to\sigma_x(\psi^\dag)^T$ and $\psi^\dag\to\psi^T\sigma_x$, magnetisation is supposed to vanish at $\mu=0$.

Evaluation of the lowest order diagram depicted in Fig.~\ref{fig:2}~(a) gives (for details, see Appendix~\ref{subsec:appendix_zeroth_order_MN})
\begin{align}
\label{eq:magnetisation}
(M_z^N)^{(0)}={}&\int_{\bm k}\frac{\Delta \vf^2}{2\vare_{\bm{k}}^3}\int_{0}^{\mu}\dd m \bigg((\vare_{\bm k}-m)f'(\vare_{\bm k}-m)\nn
&+(\vare_{\bm k}+m) f'(\vare_{\bm{k}}+m)\bigg)-(K_{xy}^{12})^{(0)}.
\end{align}
This calculation shows that the magnetisation contains a term that exactly cancels $(K_{xy}^{12})^{(0)}$ at $T=0$, together with additional terms that also vanish in this limit. As a result, the magnetisation-corrected transport coefficient $(L^{12}_{xy})_\text{tr} = K_{xy}^{12}+M_z^N$ vanishes at $T=0$, giving a well-defined $\alpha_{xy}$
\begin{align}
    \label{eq:thermoelectrix_conductivity}
    \alpha_{xy}^{(0)}={}&\frac{1}{e\bar{T}}\frac{e^2}{2h}\bigg[(-\bar{\mu})\left(1-\mathcal{F}_{-2}^{+}\right)-(\bar{\mu}+\mathcal{F}_{-1}^{-})+\nn
    &+\bar{T}\int_1^{\infty}\dd x \frac{1}{x^2}\Big(\log\fe(x)-\log\fh(x)\Big)\bigg],
\end{align}
where we introduced $\mathcal{F}_{n}^{\pm}=\int_1^\infty\dd x \,x^n(\fe(x)\pm\fh(x))$, with  $f_{e/h}(x)=1/(1+e^{(x\mp \bar{\mu})/\bar{T}})$. Here we normalise the chemical potential and the temperature by the half-gap size so that $\bar{\mu}=\mu/\Delta$ and $\bar{T}=\kb T/\Delta$.

Note that the cancellation between Eqs. \eqref{eq:K_12_0} and \eqref{eq:magnetisation} is not accidental, as both the Kubo kernel and the magnetisation are controlled by the same Berry-curvature physics of massive Dirac bands, and only their proper combination yields the physically meaningful transport response. Having established that the standard magnetisation correction fully restores a well-defined thermoelectric response in the non-interacting case, we now turn to the interacting system and show that this delicate cancellation no longer holds once electron-electron interactions are included.

\noindent

\section{First-order result}
\label{sec:first_order}

We now focus on the evaluation of the thermoelectric Hall kernel $K^{12}_{xy}$ at first-order in the electron-electron interaction. As discussed in Sec.~\ref{sec:description_of_the_model}, interactions enter the Kubo response through the usual exchange correction Fig.~\ref{fig:1}~(b), corrections to the fermionic propagators Fig.~\ref{fig:1}~(c)--(d) and through the explicit interaction-dependent contributions to the heat-current operator Fig. \ref{fig:1}~(e)--(f).

We are ignoring Hartree energy insertions and random-phase-approximation-type vertex corrections, which also arise at first-order, because these just cause a constant shift of the chemical potential.

We write the first order contribution to $K_{xy}^{12}$ as a sum of two terms:
$(K_{xy}^{12})^{(1)}=K_\mathrm{I}+K_{\mathrm{II}}$, where superscript $(1)$ indicates first order in the interaction. The term $K_\mathrm{I}$ is the sum of diagrams in Fig. \ref{fig:1} (b)--(d) and $K_{\mathrm{II}}$ is the sum of diagrams in Fig. \ref{fig:1} (e)--(f). For $K_\mathrm{I}$ we obtain
\begin{align}
 K_\mathrm{I}={}&-\mu\vf^2\bigg(\int_{\bm{k}}\dfrac{\Sigma^0(\bm{k})\Delta}{2\varepsilon_{\bm{k}}^3}\dfrac{\partial \Phi^+_{\bm{k}}}{\partial \varepsilon_{\bm{k}}}-\dfrac{\Sigma^z(\bm{k})}{2\varepsilon_{\bm{k}}^2}\dfrac{\partial \Phi_{\bm{k}}^-}{\partial \varepsilon_{\bm{k}}}\nn
 &-\Delta\int_{\bm{k}\bm{k}'}\dfrac{\vf^2(\bm{k}-\bm{k}')\cdot\bm{k}'}{4\varepsilon_{\bm{k}'}^2\varepsilon_{\bm{k}}^3}\Phi^-_{\bm{k}}\dfrac{\partial \Phi_{\bm{k}'}^-}{\partial \varepsilon_{\bm{k}'}}V_{\bm{k}-\bm{k}'}\bigg),
\end{align}
where $\Phi^{\pm}_{\bm{k}}=1-f(\varepsilon_{\bm{k}}+\mu)\pm f(\varepsilon_{\bm{k}}-\mu)$ and $\Sigma^0$ and $\Sigma^z$ are the components of the self-energy given by (see \cite{dumitriu-i.2024firstorder})
\begin{align} \Sigma_z(\bm{k})&=\dfrac{1}{2}\int_{\bm{k}'}V_{\bm{k}-\bm{k}'}
 \dfrac{\Delta}{\varepsilon_{\bm{k}}}\Phi_{\bm{k}}^-,\\
\Sigma_0(\bm{k})&=-\dfrac{1}{2}\int_{\bm{k}'}V_{\bm{k}-\bm{k}'}\Phi_{\bm{k}'}^+.
\end{align}
For $K_{\mathrm{II}}$ we arrive at
\begin{align}
    K_{\mathrm{II}}={}&-\dfrac{\Delta \vf^2}{2}\int_{\bm{k}}\bigg(\Sigma^0(\bm{k})\nn
    &+\int_{\bm{k}'}\dfrac{\vf^2(\bm{k}'-\bm{k})\cdot\bm{k}'}{4\varepsilon_{\bm{k}'}}\dfrac{\partial \Phi^+_{\bm{k}'}}{\partial \varepsilon_{\bm{k}'}}V_{\bm{k}'-\bm{k}}\bigg)\dfrac{\Phi_{\bm{k}}^-}{\varepsilon_{\bm{k}}^3}.
\end{align}

Note that, curiously, as $e^2(K^{12}_{xy})^{(0)}=-\mu\sigma^{(0)}_{xy}$, the term $K_\mathrm{I}$ is similarly related to the first-order correction to conductivity, {\it i.e.,} $e^2K_\mathrm{I}=-\mu\sigma^{(1)}_{xy}$ (see Ref. \cite{dumitriu-i.2024firstorder} for $\sigma_{xy}^{(1)}$). The existence of $K_{\mathrm{II}}$, however, prevents extending this to the whole of $(K_{xy}^{12})^{(1)}$. Also note that, like the lowest order result Eq. \eqref{eq:K_12_0}, the  correction $K_{\mathrm{I}}+K_{\mathrm{II}}$ does not vanish in the zero-temperature limit. This behaviour signals that the interacting Kubo kernel, like its non-interacting counterpart, contains equilibrium circulating currents that do not correspond to genuine transport. In the non-interacting case, these unphysical contributions are exactly cancelled by the particle magnetisation, restoring a finite and well-defined zero-temperature thermoelectric response. As we show below, however, this cancellation is no longer complete once interactions are included, even for the most local, contact-like interactions considered here. The breakdown of this cancellation, and its implications for interacting thermoelectric transport, are analysed by computing the corresponding magnetisation correction.

The sum of contributions to the right-hand side of Eq.~\eqref{eq:streda} coming from diagrams Fig. \ref{fig:2} (b)--(d) can be written as
\begin{align}\label{eq:rhs_of_streda}
\frac{\partial (M_z^N)^{(1)}}{\partial\mu}&=\vf^2\dfrac{\partial}{\partial \mu}\!\int_{\bm{k}}\!\bigg(\dfrac{\Sigma_z(\bm{k})}{2\varepsilon_{\bm{k}}}\dfrac{\partial \Phi_{\bm{k}}^+}{\partial \varepsilon_{\bm{k}}}-\dfrac{\Sigma^0(\bm{k})\Delta}{2\varepsilon_{\bm{k}}}\dfrac{\partial}{\partial \varepsilon_{\bm{k}}}\dfrac{\Phi^-_{\bm{k}}}{\varepsilon_{\bm{k}}}\bigg),\nn
\end{align}
which is surprisingly compact. Integrating this under the condition that $(M_z^N)^{(1)}$ vanishes at $\mu=0$, we obtain
\begin{equation}
    \hspace{-0.15cm}(M^N_z)^{(1)}=\vf^2\int_{\bm{k}}\bigg(\dfrac{\Sigma_z(\bm{k})}{2\varepsilon_{\bm{k}}}\dfrac{\partial \Phi_{\bm k}^+}{\partial \varepsilon_{\bm{k}}}-\dfrac{\Sigma^0(\bm{k})\Delta}{2\varepsilon_{\bm{k}}}\dfrac{\partial}{\partial \varepsilon_{\bm{k}}}\dfrac{\Phi^-_{\bm{k}}}{\varepsilon_{\bm{k}}}\bigg).
\end{equation}

The true thermoelectric response coefficient is supposed to be the sum of $K_{xy}^{12}$ and $M_z^N$, which at first order in the interaction gives
\begin{widetext}
\begin{align}\label{eq:full_correction}
(L_{xy}^{12})^{(1)}_{\text{tr}}={}&\vf^2\int_{\bm{k}}\bigg(\dfrac{\Sigma^z(\bm{k})}{2\varepsilon_{\bm{k}}^2}\bigg[\mu\dfrac{\partial \Phi_{\bm{k}}^-}{\partial\varepsilon_{\bm{k}}}+\varepsilon_{\bm{k}}\dfrac{\partial \Phi_{\bm{k}}^+}{\partial \varepsilon_{\bm{k}}}\bigg]-\dfrac{\Sigma^0(\bm{k})\Delta}{2\varepsilon_{\bm{k}}^3}\bigg[\mu\dfrac{\partial \Phi_{\bm{k}}^+}{\partial \varepsilon_{\bm{k}}}+\varepsilon_{\bm{k}}\dfrac{\partial \Phi_{\bm{k}}^-}{\partial \varepsilon_{\bm{k}}}\bigg]\nn&+\Delta\int_{\bm{k}'}\dfrac{\vf^2(\bm{k}-\bm{k}')\cdot\bm{k}'}{4\varepsilon_{\bm{k}'}^2}\bigg[\mu\dfrac{\partial \Phi_{\bm{k}'}^-}{\partial \varepsilon_{\bm{k}'}}+\dfrac{\varepsilon_{\bm{k}'}}{2}\dfrac{\partial \Phi_{\bm{k'}}^+}{\partial \varepsilon_{\bm{k}'}}\bigg]\dfrac{\Phi_{\bm{k}}^-}{\varepsilon_{\bm{k}}^3}V_{\bm{k}'-\bm{k}}\bigg).
\end{align}
\end{widetext}
This expression is the main result of this work. Note that each of the combinations 
\begin{align}\label{eq:first_combo}
&\mu\dfrac{\partial \Phi_{\bm{k}}^-}{\partial\varepsilon_{\bm{k}}}+\varepsilon_{\bm{k}}\dfrac{\partial \Phi_{\bm{k}}^+}{\partial \varepsilon_{\bm{k}}};\\
&\mu\dfrac{\partial \Phi_{\bm{k}}^+}{\partial \varepsilon_{\bm{k}}}+\varepsilon_{\bm{k}}\dfrac{\partial \Phi_{\bm{k}}^-}{\partial \varepsilon_{\bm{k}}}
\end{align}
vanishes at $T=0$. The third term in Eq.~\eqref{eq:full_correction} has a combination similar to Eq. \eqref{eq:first_combo} but not quite: the factor of $1/2$ is off. For this reason $(L_{xy}^{12})_{\text{tr}}^{(1)}$ does not vanish at $T=0$ and the approach of Ref. \cite{qianniu2011energy} fails. 
Note that if $V_{\bm{q}}$ falls off at large $\bm{q}$ no slower than a constant, Eq.~\eqref{eq:full_correction} converges. This is true, of course, modulo possible subdivergences hiding in $\Sigma^z$ and $\Sigma^0$, in which case these should be replaced by their renormalized values, with infinite parts subtracted (for renormalization, see Ref. \cite{dumitriu-i.2024firstorder}). One can also see that Eq. \eqref{eq:full_correction} does not vanish even if the interaction is taken to be a contact interaction, with $V_{\bm{q}}$ equal to a constant.  

\section{Discussion}
\label{sec:discussion}
Let us discuss possible reasons why the subtraction of magnetization currents fails at first order of perturbation theory. One possible reason could be a violation of a Ward identity. Even so, it is hard to see why this violation would occur considering that all vertex correction diagrams are explicitly finite and well defined. One, of course, could argue that they are only finite and well defined because the interaction is non-local, and by admitting a non-local interaction we violated the scaling laws of Ref. \cite{qianniu2011energy}, Eqs. \eqref{eq:scaling_law_particle} and \eqref{eq:scaling_law_energy}, which are required for the subtraction to work. However, the calculation of $(L_{xy}^{12})_{\text{tr}}$ only requires Eqs. \eqref{eq:commutators_h_n} and \eqref{eq:commutators_h_h}  to be true, which do seem to be true for any non-local potential. 

The fact that the value of the possible missing term depends on the existence of the Fermi surface is somewhat puzzling, although such dependence is in fact necessary, since in absence of the Fermi surface Eq. \eqref{eq:full_correction} vanishes at $T=0$. The reason this is puzzling is that whatever violation of either the Ward identity or Eqs.~\eqref{eq:scaling_law_particle} and \eqref{eq:scaling_law_energy}
is inadvertently introduced into the calculation must come from the small distance scale and hence must be largely insensitive to the infrared physics implied by existence or absence of the Fermi surface.

Another possibility is that an $(L_{xy}^{12})_{\text{tr}}$ that does not vanish at $T=0$ is a genuine feature of Hall thermoelectric response in the system with interactions. This would mean that interactions introduce a non-analytic behavior into $\alpha_{xy}$, whose real form can only be uncovered by summing up an infinite number of leading diagrams in each order of perturbation theory. We, however, deem this possibility very unlikely, as it does not seem physically intuitive. 

\section{Conclusion}
In this work we calculated the Hall thermoelectric response of massive Dirac fermions in two space dimensions to first order in perturbation theory. We calculated both  the Kubo susceptibility and the particle magnetization. We found that the expected cancellation between them at $T=0$ does not occur.
A possible reason why this happens could be a subtle violation of locality that leads to a violation of a Ward identity or the scaling laws, Eqs. \eqref{eq:scaling_law_particle} and \eqref{eq:scaling_law_energy}. 

\label{sec:conclusion}
\begin{acknowledgments}
A.D.D.-I. acknowledges support from the Engineering and Physical Sciences Research Council, Grant No. EP/T517823/1. A.P. and A.E.K. acknowledge support from the Leverhulme Trust under the Grant Agreement No. RPG-2023-253. The authors also acknowledge support from the European Commission under the EU Horizon 2020 MSCA-RISE-2019 programme (Project No. 873028 HYDROTRONICS).
\end{acknowledgments}

\allowdisplaybreaks
\appendix
\renewcommand{\thesubsection}{\thesection.\arabic{subsection}}
\makeatletter
\renewcommand{\p@subsection}{}
\makeatother
\section{\label{sec:appendix_zeroth_order}Zeroth-order result}

In this appendix, we cover the technical details of the non-interacting calculation underlying the results presented in the main text based on Eqs. \eqref{eq:kubo}--\eqref{eq:streda}. To calculate $(\Pi^R)_{ij}^{12}(\omega)$, we first calculate imaginary time susceptibility
\begin{equation}\label{eq:matsubara_bubble}
    (\Pi^M)^{12}_{ij}(i\omega_m)=-\dfrac{1}{V}\int_0^\beta d\tau e^{i\omega_m\tau}\Big\langle\text{T}_{\tau}(\hat{\bm{j}}_0^N)_i(\tau)(\hat{\bm{j}}_0^Q)_j\Big\rangle,
\end{equation}
where $\text{T}_\tau$ is time-ordering in imaginary time, which we then analytically continue to real frequencies. Here $\hat{\bm{j}}_0^N(\tau)=e^{\mathcal{H}\tau}\hat{\bm{j}}_0^N e^{-\mathcal{H}\tau}$ and subscript $0$ stands for vanishing momentum. 
Eq. \eqref{eq:matsubara_bubble} can be evaluated as a sum of Feynman diagrams \cite{LL}.

On the other hand, correlator $\Big\langle \hat{\bm{j}}^N_{\bm{q}}; \hat{n}_{-\bm{q}}\Big\rangle V^{-1}$ can also be represented as an imaginary time susceptibility at vanishing frequency
\begin{equation}\label{eq:static_susceptibility}
    \beta V^{-1}\Big\langle \hat{\bm{j}}_{\bm{q}}^N;\hat{n}_{-\bm{q}}\Big\rangle = V^{-1}\int_0^\beta d\tau\Big\langle \text{T}_\tau \hat{\bm{j}}_{\bm{q}}(\tau) \hat{n}_{-\bm{q}}(0)\Big\rangle
\end{equation}
and thus can also be calculated as a sum of Feynman diagrams. 

In the non-interacting limit, the diagrammatic expansion of either Eq. \eqref{eq:matsubara_bubble} or \eqref{eq:static_susceptibility} reduces to a single bubble diagram, with two single-particle Green’s functions attached to the appropriate operator vertices, as shown for example in Figs.~\ref{fig:1}(a) or~\ref{fig:2}(a). At finite external momentum $\bm q$ and frequency $i\omega_m$ for any two operators $\hat{A}_{\bm{q}}$ and $\hat{B}_{-\bm{q}}$, the noninteracting bubble has the form
\begin{align}
\label{eq:appendix_correlation_function}
&\int_0^\beta d\tau e^{i\omega_m\tau}\Big\langle  \text{T}_\tau \hat{A}_{\bm{q}}(\tau) \hat{B}_{-\bm{q}}(0) \Big\rangle\nn
={}&-\sum_{n}\int \dd \vare_1 \dd\vare_2 \frac{\tr \big[\hat{A}_{\bm{q}}\delta(\vare_1-\hat{H}_0)\hat{B}_{-\bm{q}}\delta(\varepsilon_{2}-\hat{H}_0)\big]}{\big(i(\varepsilon_n+\omega_m)+\mu-\vare_1\big)\big(i\varepsilon_{n}+\mu-\vare_2\big)}\nn
={}&\int \dd\vare_1 \dd\vare_2  \frac{f_e(\vare_2)-f_e(\vare_1)}{\vare_2-\vare_1+i\omega_m}\times\nn
&\times\tr \big[\hat{A}_{\bm{q}}\delta(\vare_1-\hat{H}_0)\hat{B}_{-\bm q}\delta(\vare_2-\hat{H}_0)\big],
\end{align}
where $f_e(\varepsilon)=(e^{(\varepsilon-\mu)/T}+1)^{-1}$ is the Fermi--Dirac distribution for electrons. Note that a slight abuse of notation happened here: in the top line of this equation the operators are many-body but in the subsequent lines they are one-body.

\subsection{\label{subsec:appendix_zeroth_order_kubo}Kubo part}

Using Eq. \eqref{eq:appendix_correlation_function} for $(\Pi^M)_{ij}^{12}(i\omega_m)$ we obtain
\begin{align}
\label{eq:appendix_kubo_pi12}
&(\Pi^M)_{xy}^{12}(i\omega_m)=-\dfrac{1}{V}\int \dd\vare_1 \dd\vare_2\frac{f_e(\vare_2)-f_e(\vare_1)}{\left(\vare_2-\vare_1+i\omega_m\right)}\nn
&\times \tr\left[\delta(\vare_1-\hat{H}_0)(\hat{\bm{j}}^N_0)_x\delta(\vare_2-\hat{H}_0)(\hat{\bm{j}}^Q_0)_{y}\right]\nn
={}& -\dfrac{1}{V}\int \dd\vare_1 \dd\vare_2\frac{f_e(\vare_2)-f_e(\vare_1)}{\left(\vare_2-\vare_1+i\omega_m\right)}\frac{\vare_1+\vare_2-2\mu}{2}\nn
&\times \tr\left[\delta(\vare_1-\hat{H}_0)\hat{v}_{x}\delta(\vare_2-\hat{H}_0)\hat{v}_{y}\right],
\end{align}
 where we used definitions Eqs. \eqref{eq:particle_current}, \eqref{eq:energy_current} and that $\hat{\bm{j}}^Q=\hat{\bm{j}}^E-\mu\hat{\bm{j}}^N$ and  the identity $\hat{H}_0\delta(\vare-\hat{H}_0)=\vare\delta(\vare-\hat{H}_0)$. The trace can be further evaluated by inserting the resolution of identity in the single-particle eigenbasis. This gives a sum over two band indices $\lambda$, $\lambda'$ and momentum $\bm{k}$ and allows us to rewrite the product inside the trace in terms of projection operators $\Pi_{\bm k}^{\lambda}$ onto the eigenstates. Note that since $\tr\left[\Pi^{\lambda}_{\bm k}\hat{v}_{x}\Pi^{\lambda\pr}_{ \bm k}\hat{v}_{y}\right]=(1-\delta_{\lambda\lambda\pr})\lambda(i\vf^2\Delta)/\vare_{\bm k}$, only interband terms contribute to the sum so that we obtain
\begin{align}
(\Pi^M)^{12}_{xy}
&=\mu\int_{\bm k}\frac{f_e(\vare_{\bm{k}})-f_e(-\vare_{\bm{k}})}{(2\vare_{\bm{k}})^2-(i\omega_m)^2}\tr\Big(\Pi_{\bm k}^+\hat{v}_x\Pi_{\bm{k}}^-\hat{v}_y\Big)(2i\omega_m)\nn
\end{align}
where we used the shorthand notation $\int_{\bm k}=\int \dd\bm k/{(2\pi)^2}$. Analytically continuing to real frequencies and using formula \eqref{eq:handy_formula}, we obtain
\begin{align}\label{eq:appendix_K_12_final_form}
    (K_{xy}^{12})^{(0)}&=i\mu\int_{\bm{k}}\dfrac{f_e(\varepsilon_{\bm k})-f_e(-\varepsilon_{\bm{k}})}{2\varepsilon_{\bm{k}}^2}\tr\Big(\Pi_{\bm k}^+\hat{v}_x\Pi_{\bm{k}}^-\hat{v}_y\Big)\nn    
    &=-\mu\Delta \vf^2\int_{\bm{k}}\dfrac{f_e(\varepsilon_{\bm{k}})-f_e(-\varepsilon_{\bm{k}})}{2\varepsilon_{\bm{k}}^3}\nn
    &=\dfrac{\bar{\mu}}{4\pi}(1-\mathcal{F}^+_{-2}),
\end{align}
where $\bar{\mu}=\mu/\Delta$ and $\mathcal{F}_n^{\pm}=\int_1^\infty dx x^n(f_e(x)\pm f_h(x))$, with $f_{e/h}(x)=(e^{(x\mp \bar{\mu})/\bar{T}}+1)^{-1}$.

We observe that the thermoelectric kernel satisfies $e^2K^{12}_{xy} = (-\mu) \sigma_{xy}$, where $\sigma_{xy}$ is the anomalous Hall conductivity. This can be traced back to the fact that only inter-band terms, in which $\varepsilon_1$ and $\varepsilon_2$ lie in the opposite bands (so that $\varepsilon_1+\varepsilon_2=0$), contribute to \eqref{eq:appendix_kubo_pi12}.

In the limit of zero temperature, $1-\mathcal{F}_{-2}^{+} \to\Delta\,\text{sgn}(\mu)/\mu$ for $|\mu|>\Delta$, while $\mathcal{F}_{-2}^{+}\to0$ when the chemical potential lies in the gap. Substituting these limits into \eqref{eq:appendix_K_12_final_form} we see that $(K_{xy}^{12})^{(0)}$ stays finite as  $T \to 0$. This means that $\alpha_{xy}$ (see Eq. \eqref{eq:alpha}) diverges in the same limit and we obtain an infinite response. This divergence reflects the presence of equilibrium circulating currents rather than genuine transport. This signals the necessity of including the magnetisation correction.

\subsection{\label{subsec:appendix_zeroth_order_MN}Magnetisation part}

Using Eqs. \eqref{eq:static_susceptibility} and \eqref{eq:appendix_correlation_function} to calculate the static current-density susceptibility to first order in $\bm{q}$, we get
\begin{align}
\label{eq:appendix_current_density_finite_momentum_non-interacting}
\beta V^{-1}&\Big\langle  \hat{n}_{-\bm q}; \hat{\bm{j}}^N_{\bm{q}} \Big\rangle= \dfrac{1}{V}\int \dd\vare_1 \dd\vare_2\frac{f_e(\vare_1)-f_e(\vare_2)}{\vare_1-\vare_2}\nn
&\times\tr\Big[\delta(\vare_1-H)\dfrac{\hat{n}_{\bm{q}}\hat{\bm{v}}+\hat{\bm{v}}\hat{n}_{\bm{q}}}{2}\delta(\vare_2-H)n_{-\bm{q}}\Big]\nn
={}&\frac{1}{V}\int \dd \vare_1 \dd\vare_2\frac{1}{\vare_1-\vare_2}\nn
&\times \left(\frac{f_e(\vare_1)-f_e(\vare_2)}{\vare_1-\vare_2}-\dfrac{f'(\varepsilon_2)+f'(\varepsilon_1)}{2}\right)\nn
&\times\tr\left[\delta(\vare_1-H)\big(\bm{q}\cdot \hat{\bm{v}}\big)\delta(\vare_2-H)\hat{\bm{v}}\right],
\end{align}
where terms of order $\bm{q}^2$ and higher have been discarded.

Taking the curl of this correlator above, we find that Eq.~\eqref{eq:streda} gives
\begin{align}
\label{eq:appendix_partial_mu_M_N_non-interacting}
&\hspace{-0.5cm}\frac{\partial (\bm M^N)^{(0)}}{\partial \mu}=-\frac{\beta V^{-1}}{2i}\lim_{\bm q\to0}\bm \nabla_{\bm q}\times \left\langle n_{-\bm q}; \bm j^N_{\bm q}\right\rangle\nn
={}&-\frac{1}{2i}\vare_{ijk}\bm e_i V^{-1}\int \dd \vare_1 \dd\vare_2\frac{1}{\vare_1-\vare_2}\nn
&\times \left(\frac{f_e(\vare_1)-f_e(\vare_2)}{\vare_1-\vare_2}-\frac{f_e\pr(\vare_1)+f_e\pr(\vare_2)}{2}\right)\nn
&\times\tr\left[\delta(\vare_1-H)\hat{v}_j\delta(\vare_2-H)\hat{v}_k\right]\nn
={}&-\frac{1}{2i}\bm e_z \int_{\bm k} \frac{1}{\vare_{\bm k}}\tr\left[\Pi^{+}_{ \bm k}\hat{v}_{x}\Pi^{-}_{\bm k}\hat{v}_{y}\right]\nn
&\times\left(\frac{f_e(\vare_{\bm k})-f_e(-\vare_{\bm k})}{\vare_{\bm k}}-f_e\pr(\vare_{\bm k})-f_e\pr(-\vare_{\bm k})\right).\hspace{-0.75cm}
\end{align}
This equation has to be integrated with respect to $\mu$ to produce $(M^N_z)^{(0)}$. As the initial value surface we pick $\mu=0$, where $M^N_z$ has to vanish due to particle-hole symmetry. (Note that the right-hand side of Eq. \eqref{eq:appendix_partial_mu_M_N_non-interacting} is even in $\mu$.) On the first term in the last line of Eq. \eqref{eq:appendix_partial_mu_M_N_non-interacting}, we will use the identity
\begin{align}
    \int_0^\mu &d\mu'\Big(f(\varepsilon_{\bm k}-\mu')-f(-\varepsilon_{\bm{k}}-\mu')\Big)\nn={}&\mu \Big(f(\varepsilon_{\bm k}-\mu)-f(-\varepsilon_{\bm{k}}-\mu)\Big)\nn&-\int_0^\mu d\mu'~\mu'~\dfrac{\partial}{\partial \mu'}\Big(f(\varepsilon_{\bm{k}}-\mu')-f(-\varepsilon_{\bm{k}}-\mu')\Big).
\end{align}
On the second term, we will use the fact that 
\begin{equation}
f_e'(\varepsilon_{\bm{k}})+f_e'(-\varepsilon_{\bm{k}})=-\dfrac{\partial}{\partial \mu}\Big(f(\varepsilon_{\bm{k}}-\mu)+f(-\varepsilon_{\bm k}-\mu)\Big).
\end{equation}
Using these two, we can integrate Eq. \eqref{eq:appendix_partial_mu_M_N_non-interacting} to obtain $(M_z^N)^{(0)}$ in the form
\begin{widetext}
\begin{equation}
\label{eq:appendix_M_N_non-interacting}
\hspace{-0.2cm}(M^N_z)^{(0)}(\mu)=-(K_{xy}^{12})^{(0)}-i \int_{\bm k}\tr\left[\Pi^{+}_{ \bm k}\hat{v}_{x}\Pi^{-}_{\bm k}\hat{v}_{y}\right]\frac{1}{2\vare_{\bm k}^2}\int_{0}^{\mu}\dd m \left(\!\big(m-\varepsilon_{\bm{k}}\big) \frac{\partial f(\vare_{\bm k}-m)}{\partial m}-\big(m+\varepsilon_{\bm{k}}\big) \frac{\partial f(-\vare_{\bm k}-m)}{\partial m}\!\right),
\end{equation}
\end{widetext}
We therefore find that $(M^N_z)^{(0)}$ can be written as the negative of the Kubo kernel plus an additional term that vanishes identically in the zero-temperature limit.  As a result, after the \enquote{magnetisation subtraction}, the transport kernel
$(L^{12}_{xy})^{(0)}_{\mathrm{tr}} = (K^{12}_{xy})^{(0)} + (M^N_z)^{(0)}$ given by
\begin{widetext}
\begin{equation}
    (L_{xy}^{12}\big)_{\text{tr}}^{(0)}=-i\int_{k}\dfrac{\tr\Big(\Pi_{\bm{k}}^+\hat{v}_x\Pi_{\bm{k}}^{-}\hat{v}_y\Big)}{2\varepsilon_{\bm{k}}^2}\int_0^\mu \dd m\Big(\big(m-\varepsilon_{\bm{k}}\big)\dfrac{\partial f(\varepsilon_{\bm{k}}-m)}{\partial m}-\big(m+\varepsilon_{\bm{k}}\big)\dfrac{\partial f(-\varepsilon_{\bm{k}}-m)}{\partial m}\Big)
\end{equation}
\end{widetext}
goes to zero as $T \to 0$. Note that with the anomalous Hall conductivity given at $T=0$ to zeroth order by
\begin{equation}
    \sigma_{xy}^{(0)}(\mu)=ie^2\int_{\bm{k}}\dfrac{\tr\big(\Pi_{\bm{k}}^+\hat{v}_x\Pi_{\bm{k}}^-\hat{v}_y\big)}{2\varepsilon_{\bm{k}}^2}\Big(\theta(\mu-\varepsilon_{\bm{k}})-\theta(\mu+\varepsilon_{\bm{k}})\Big),
\end{equation}
coefficient $\big(L_{xy}^{12}\big)_{\text{tr}}$ can be written in the form
\begin{equation}
    (L^{12}_{xy})_{\text{tr}}=\dfrac{1}{e^2}\int d\varepsilon \sigma_{xy}^{(0)}(\varepsilon)(\varepsilon-\mu)\dfrac{\partial f(\varepsilon-\mu)}{\partial \mu},
\end{equation}
which is nothing but the Mott relation for anomalous transport (see Ref. \cite{xiao2006berryphase}).

Even though the expression in~\eqref{eq:appendix_M_N_non-interacting} better shows the cancellation with the Kubo term, the integral with respect to $\bm{k}$ in Eq.~\eqref{eq:appendix_partial_mu_M_N_non-interacting} can, in fact, be easily evaluated to give 
\begin{equation}
\label{eq:appendix_partial_mu_M_N_non-interacting_simplified}
    \frac{\partial \bm M^N}{\partial \mu} =-\bm e_z\frac{\fe(\Delta)-\fe(-\Delta)}{4\pi}.
\end{equation}

\section{\label{sec:appendix_first_order}First-order interaction corrections}

In this appendix we present the detailed derivation of the first-order interaction corrections to the thermoelectric Kubo kernel and to the particle magnetisation. We aim to do the calculations for a short-ranged, contact-like two-body interaction, whose $V_{\bm{k}}$ is effectively momentum-independent at small $\bm{k}$ (large length scales), as expected for an overscreened Coulomb interaction. At large $\bm{k}$ (tiniest length scales), however, an explicit momentum dependence of $V_{\bm{k}}$ is retained to regularise the ultraviolet divergences.

Exchange and self-energy diagrams contributing to  $(K_{xy}^{12})^{(1)}$ and $(M_{z}^N)^{(1)}$ (see Fig.~\ref{fig:1}~(b)--(d) and Fig.~\ref{fig:2}~(b)--(d), respectively) only differ in what operators are put in external vertices and the limits in which they are evaluated.  The transport Kubo kernel is obtained by studying the small-$\omega$ behaviour of particle current--heat current correlator at $\bm{q}=0$ (see Eq.~\eqref{eq:handy_formula}) whereas magnetisation follows from the small-$\bm q$ behaviour of the static ($\omega=0$) density--current correlator via the St\v{r}eda formula, see Eq.~\eqref{eq:streda}.  For clarity and to avoid duplication, we first derive the general expressions for each diagram at finite external momentum and frequency, and then apply to them the Kubo and magnetisation limits in separate subsections. The interaction-induced vertex correction to the heat current, which contributes only to the Kubo response, Fig.~\ref{fig:1}~(e)--(f), is treated separately in the final part of this section.

In the calculations that follow, both the Kubo kernel and the particle magnetisation reduce to traces over products of band projectors and velocity operators. This reflects the fact that, after resolving the fermionic degrees of freedom in the band basis of the massive Dirac Hamiltonian, all first-order response functions are expressed in terms of closed fermionic loops with appropriate operator insertions. It is therefore convenient to identify the basic trace structures that recur throughout the calculation:
\begin{widetext}
\begin{align}
\label{eq:trace_1}
\tr\!\left[\Pi_{\bm k}^{\lambda_1}\hat{v}_x\Pi_{\bm k}^{\lambda_{2}}\Pi_{\bm k\pr}^{\lambda_3}\hat{v}_y\Pi_{\bm k\pr}^{\lambda_{4}}\right]
&=\frac{i\vf^2\Delta}{8}\left[\frac{\lambda_1-\lambda_{2}}{\vare_{\bm k}}+\frac{\lambda_{4}-\lambda_3}{\vare_{\bm{k}\pr}}+\frac{\vf^2 \bm k \cdot \bm k\pr +\Delta^2}{\vare_{\bm k} \vare_{\bm{k}\pr}}\left(\frac{\lambda_1\lambda_{2}(\lambda_3-\lambda_{4})}{\vare_{\bm k}}+\frac{\lambda_{4}\lambda_3(\lambda_{2}-\lambda_1)}{\vare_{\bm{k}\pr}}\right)\!\right]\!,\\\label{eq:trace_2}
\tr\!\left[\Pi_{\bm k}^{\lambda_1}\hat{v}_x\Pi_{\bm k}^{\lambda_{2}}\hat{v}_y\Pi_{\bm k}^{\lambda_3}\Pi_{\bm k\pr}^{\lambda_{4}}\right]
&=\frac{i\vf^2 \Delta}{8}\left[\frac{(\lambda_1-\lambda_{2})(1+\lambda_1\lambda_3)}{\vare_{\bm k}}+\frac{\lambda_{4}(1-\lambda_1\lambda_3)}{\vare_{\bm{k}\pr}}+\frac{\vf^2 \bm k \cdot \bm k\pr +\Delta^2}{\vare_{\bm k} \vare_{\bm{k}\pr}}\frac{\lambda_1\lambda_{4}}{\vare_{\bm k}}\left(2\lambda_3-\lambda_{2}(1+\lambda_1\lambda_3)\right)\!\right]\!,
\end{align}
\end{widetext}
where the terms that vanish upon angular integration have been discarded. All other trace expressions encountered below follow from these by cyclicity of the trace and the antisymmetry of both the Hall response and the magnetisation under exchange of spatial indices $x\leftrightarrow y$.

\subsection{\label{subsec:appendix_exchange}Exchange diagram}

Contribution of the exchange diagram to Matsubara susceptibility of two densities $\hat{A}_{\bm{q}}$ and $\hat{B}_{\bm{-q}}$ at momentum $\bm{q}$ and frequency $i\omega_m $ equals (see Ref. \cite{giuliani2005quantum} for Feynman rules)
\begin{align}
\label{eq:appendix_AB_EX}
&-\int_0^\beta d\tau e^{i\omega_m\tau} \left\langle  \text{T}_\tau \hat{A}_{\bm{q}}(\tau) \hat{B}_{-\bm{q}}(0) \right\rangle^{\text{(EX)}}V^{-1}\nn
={}&-\dfrac{1}{V}\int_{\bm l}V_{\bm l}\frac{1}{\beta^2}\sum_{n,n\pr}\int \dd\vare_1 \dd\vare_2 \dd\vare_3 \dd\vare_4 \nn
&\times \tr \left[\hat{A}_{\bm q}\frac{\delta(\vare_1-\hat{H}_0)}{i(\varepsilon_n+\omega_m)+\mu-\vare_1}\hat{n}_{-\bm l}\frac{\delta(\vare_2-\hat{H}_0)}{i(\varepsilon_{n\pr}+\omega_m)+\mu-\vare_2}\right.\nn
&\left.\times \hat{B}_{-\bm q}\frac{\delta(\vare_3-\hat{H}_0)}{i\varepsilon_{n\pr}+\mu-\vare_3}\hat{n}_{\bm l}\frac{\delta(\vare_4-\hat{H}_0)}{i\varepsilon_n+\mu-\vare_4}\right]\nn
={}& -\dfrac{1}{V}\int_{\bm l}V_{\bm l}\int \dd\vare_1 \dd\vare_2 \dd\vare_3 \dd\vare_4 \frac{\fe(\vare_1)-\fe(\vare_4)}{\vare_1-\vare_4-i\omega_m}\nn
&\times\frac{\fe(\vare_2)-\fe(\vare_3)}{\vare_2-\vare_3-i\omega_m}
\tr \left[\hat{A}_{\bm q}\delta(\vare_1-\hat{H}_0)\hat{n}_{-\bm l}\delta(\vare_2-\hat{H}_0)\right.\nn
&\left.\times \hat{B}_{-\bm q}\delta(\vare_3-\hat{H}_0)\hat{n}_{\bm l}\delta(\vare_4-\hat{H}_0)\right],
\end{align}
where $\bm l$ denotes the momentum flowing along the interaction line and we used spectral decomposition of the one-particle Green function to sum on the fermionic Matsubara frequencies. (Note the inevitable abuse of notation: operators $\hat{A}_{\bm{q}}$ and $\hat{B}_{-\bm{q}}$ are many-body in the top line and one-body in all the subsequent lines.)

Choosing $\hat{A}_{\bm{q}}=(\hat{\bm{j}}^{N}_{\bm{q}})_x$ and $\hat{B}_{-\bm{q}}= (\hat{\bm{j}}^Q_{-\bm {q}})_y$, with $\hat{\bm{j}}^Q=\hat{\bm{j}}_E-\mu \hat{\bm{j}}^N$, setting $\bm{q}$ to zero (but keeping $i\omega_m$ finite) and using definitions Eqs. \eqref{eq:particle_current} and \eqref{eq:energy_current}, we obtain

\begin{align}
\label{eq:appendix_jNjQ_EX}
-\int_0^\beta &d\tau e^{i\omega_m\tau}\left\langle  \text{T}_\tau(\hat{\bm{j}}^N_0)_x(\tau)(\hat{\bm{j}}^{Q}_0)_{y}(0)\right\rangle^{\text{(EX)}}V^{-1}\nn
={}&-\sum_{\substack{\lambda_1, \lambda_2 \\ \lambda_3, \lambda_4}}\int_{\bm k}\int_{\bm k\pr}V_{\bm k-{\bm k}\pr} \bigg(\frac{\fe(\vare_{\lambda_4 {\bm k}})-\fe(\vare_{\lambda_1 {\bm k}})}{\vare_{\lambda_4 {\bm k}}-\vare_{\lambda_1 {\bm k}}-i\omega_m}\nn&{}\times\frac{\fe(\vare_{\lambda_3 {\bm k}\pr})-\fe(\vare_{\lambda_2 {\bm k}\pr})}{\vare_{\lambda_3 {\bm k}\pr}-\vare_{\lambda_2 {\bm k}\pr}-i\omega_m}\bigg)\frac{\vare_{\lambda_2 {\bm {\bm k}'}}+\vare_{\lambda_3 {\bm{ k}'}}-2\mu}{2}\nn
&\times{}\tr\left[\Pi^{\lambda_1}_{ \bm k}\hat{v}_{x}\Pi^{\lambda_4}_{\bm k}\Pi^{\lambda_3}_{ {\bm k}\pr}\hat{v}_y\Pi^{\lambda_2}_{{\bm k}\pr}\right],
\end{align}
where we replaced the delta-functions by their spectral representations in Eq. \eqref{eq:appendix_AB_EX} and integrated over $\varepsilon_{1,2,3,4}$.
After performing the analytical continuation, employing formula \eqref{eq:handy_formula} and evaluating the trace using Eq. \eqref{eq:trace_1}, we obtain
\begin{align}
\label{eq:appendix_K12_EX}
K_{xy}^{12\text{(EX)}}&=
\mu\frac{\vf^2 \Delta}{4}\int_{\bm k,\bm k\pr}V_{\bm k-\bm k\pr}\frac{\vare_{\bm k}^2+\vf^2\bm k\cdot{\bm k}\pr+\Delta^2}{\vare_{\bm k}^3\vare_{\bm{k}\pr}^3}\Phi_{\bm k}^{-}\Phi_{\bm{k}\pr}^{-}
\end{align}
where $\Phi^{\pm}_{\bm k}=\pm \fe(\vare_{\bm k})+\fe(-\vare_{\bm k})$. Note that, curiously, because it is only the terms with $\lambda_2=-\lambda_3$ that contribute to the sum in Eq. \eqref{eq:appendix_K12_EX}, exchange contribution to $K_{xy}^{12}$ is $(-\mu/e^2)$ times the exchange contribution to anomalous Hall conductivity.

In the magnetisation calculation we choose $\hat{A}_{\bm{q}}=\hat{\bm{j}}^N_{\bm{q}}$ and $\hat{B}_{-\bm{q}}=\hat{n}_{-\bm{q}}$, which we substitute into Eq.~\eqref{eq:appendix_AB_EX} and set $i\omega_m$ to zero. Then we can use the fact that $\hat{n}_{\bm{q}}\hat{H}_0\hat{n}_{-\bm{q}}=\hat{H}_0+\bm{q}\cdot\hat{\bm{v}}+O(\bm{q}^2)$ and that
\begin{align}    \delta(\varepsilon-\hat{H}_0-\bm{q}\cdot\hat{\bm{v}})={}&\int \dd\varepsilon_{1}\dd\varepsilon_{2}\dfrac{\delta(\varepsilon-\varepsilon_{2})-\delta(\varepsilon-\varepsilon_1)}{\varepsilon_2-\varepsilon_1}\nn
&\times\delta(\varepsilon_1-\hat{H}_0)\bm{q}\cdot\hat{\bm{v}}\delta(\varepsilon_2-\hat{H}_0)
\end{align}
to expand the right-hand side of Eq.~\eqref{eq:appendix_AB_EX} to first order in $\bm{q}$ and take the curl according to Eq.~\eqref{eq:streda}\footnote{Note that in Eq. \eqref{eq:appendix_AB_EX} it may actually be easier to  expand to first order in $\bm{q}$ first and then calculate the Matsubara sum.}.

Afterwards, replacing the delta-functions by their spectral decompositions and calculating the traces with Eqs.~\eqref{eq:trace_1}--\eqref{eq:trace_2}, we arrive at
\begin{align}
    &\frac{\partial M_z^{N\text{(EX)}}}{\partial \mu}=- \frac{\vf^2\Delta}{4} \int_{\bm k,\bm k\pr}V_{\bm k-\bm k\pr}\bigg[\frac{\partial}{\partial \vare_{\bm {k}\pr}}\left(\frac{\Phi_{\bm{k}\pr}^{-}}{\vare_{\bm{k}\pr}}\right)\nn&{}\times\left\{\frac{\vf^2 \bm k \cdot \bm k\pr+\Delta^2}{2 \vare_{\bm k} \vare_{\bm{k}\pr}}\frac{\partial}{\partial \vare_{\bm k}}\left(\frac{\Phi_{\bm k}^{-}}{\vare_{\bm k}}\right)\right.-\left.\frac{1}{2 \vare_{\bm k} \vare_{\bm{k}\pr}}\frac{\partial}{\partial \vare_{\bm k}}\left(\vare_{\bm k} \Phi_{\bm k}^{-}\right)\right\}\nn
    &{}-\dfrac{1}{\varepsilon_{\bm{k}}}\dfrac{\partial \Phi_{\bm{k}'}^-}{\partial \varepsilon_{\bm{k}'}}\dfrac{\partial}{\partial \varepsilon_{\bm{k}}}\bigg(\dfrac{\Phi_{\bm{k}}^-}{\varepsilon_{\bm{k}}}\bigg)-\dfrac{1}{\varepsilon_{\bm{k}'}\varepsilon_{\bm{k}}}\dfrac{\partial \Phi_{\bm{k}'}^+}{\partial \varepsilon_{\bm{k}'}}\dfrac{\partial \Phi^+_{\bm{k}}}{\partial \varepsilon_{\bm{k}}}
       \bigg].
\end{align}

\subsection{\label{subsec:appendix_self_energy}Self-energy diagram}

The self-energy contribution arises from self-energy insertions on the fermionic propagators and corresponds diagrammatically to the two diagrams shown in Figs.~\ref{fig:1}~(c)--(d) and~\ref{fig:2}~(c)--(d). Rather than evaluating these diagrams, it is more practical to evaluate diagrams in Figs.~\ref{fig:1}~(a) and \ref{fig:2}~(a) instead but with Green functions that already include the self-energy, {\it i.e.,} with momentum-dependent effective chemical potential, Fermi velocity and mass. At first order in the interaction, this procedure exactly reproduces the two self-energy diagrams and avoids unnecessary duplication of algebra.

More concretely, given the self-energy $\Sigma(\bm{k})=\Sigma^0(\bm{k})+\bm{\sigma}\cdot\bm{\Sigma}^\parallel(\bm{k})+\sigma_z\Sigma^z(\bm{k})$ (note that $\Sigma$ does not depend on $i\omega_m$), we do the replacement
\begin{align}\label{eq:first_replacement}
    &\mu\rightarrow\mu-\Sigma^0(\bm{k})\\\label{eq:second_replacement}
    &\vf\rightarrow\vf+\dfrac{\bm{\Sigma}^{\parallel}(\bm{k})\cdot{\bm{k}}}{\bm{k}^2}\\\label{eq:third_replacement}
    &\Delta\rightarrow\Delta+\Sigma^z(\bm{k})
\end{align}
in the noninteracting diagrams and then expand to first order in $\Sigma$. The self-energy components are given by (see Ref. \cite{dumitriu-i.2024firstorder})
\begin{align}
&\Sigma^0=-\dfrac{1}{2}\int_{\bm{k}'}V_{\bm{k}-\bm{k}'}\Phi^+_{\bm{k}'};
\\
&\Sigma^{z}=\dfrac{1}{2}\int_{\bm{k}'}V_{\bm{k}-\bm{k}'}\dfrac{\Delta}{\varepsilon_{\bm{k}'}}\Phi_{\bm{k}'}^-;\\
&\bm{\Sigma}^{\parallel}=\dfrac{1}{2}\int_{\bm{k}'}V_{\bm{k}-\bm{k}'}\dfrac{\vf\bm{k}'}{\varepsilon_{\bm{k}'}}\Phi^-_{\bm{k}'}.
\end{align}
Following these steps, we can obtain the contribution of diagrams in Figs.~\ref{fig:1}~(c)--(d) to $K_{xy}^{12}$ by doing replacements \eqref{eq:first_replacement}--\eqref{eq:third_replacement} in result \eqref{eq:appendix_K_12_final_form}, which gives

\begin{widetext}
\begin{equation}
\label{eq:appendix_K12_SE}
K_{xy}^{12\text{(SE)}}
=(-\mu)\frac{\vf\Delta}{2}\!\int_{\bm k}\!\left\{\Sigma^0(\bm{k})\left[\frac{\vf}{\vare_{\bm k}^{3}}\frac{\partial \Phi^{+}_{\bm k}}{\partial\vare_{\bm k}}\right]-\bm{\Sigma}^\parallel(\bm{k})\cdot \left[\bm \nabla_{\bm k}\left(\frac{\Phi^{-}_{\bm k}}{\vare_{\bm k}^{3}}\right)\right]+\Sigma^z(\bm{k})\left[\frac{\vf}{\Delta}\bm \nabla_{\bm k}\cdot \left( \bm k \frac{\Phi^{-}_{\bm k}}{\vare_{\bm k}^{3}}\right)-\frac{\vf}{\vare_{\bm k}^{2}\Delta}\frac{\partial \Phi^{-}_{\bm k}}{\partial \vare_{\bm k}}\right]\!\right\}.
\end{equation}
\end{widetext}
Importantly, the shifts \eqref{eq:first_replacement}--\eqref{eq:third_replacement}  apply only to the parameters entering the Green’s functions, while $\mu$ and $v_F$ entering the vertices remain unchanged. This reflects the fact that the self-energy renormalises the fermionic propagators but does not generate vertex corrections.

Note again that Eq.~\eqref{eq:appendix_K12_SE} obeys the $e^{2}K_{xy}^{12}=-\mu\sigma_{xy}$ relation like the non-interacting result \eqref{eq:appendix_K_12_final_form} and the exchange contribution \eqref{eq:appendix_K12_EX}. 

The particle magnetisation that comes from diagrams in Fig.~\ref{fig:2}~(c)--(d) can be obtained from the non-interacting calculation by replacing the bare parameters  by their interaction-corrected counterparts according to \eqref{eq:first_replacement}--\eqref{eq:third_replacement}. After applying this change  to Eq.~\eqref{eq:appendix_partial_mu_M_N_non-interacting}, and expanding to first order in the interaction strength we obtain the following first-order self-energy correction to the magnetisation:
\begin{align}
    &\dfrac{\partial M_z^{N(\mathrm{SE})}}{\partial \mu}=\vf^2\int_{\bm{k}}\bigg(-\dfrac{\Delta\Sigma^0(\bm{k})}{2\varepsilon_{\bm{k}}}\dfrac{\partial}{\partial \varepsilon_{\bm{k}}}\dfrac{1}{\varepsilon_{\bm{k}}}\dfrac{\partial \Phi^+_{{\bm{k}}}}{\partial \varepsilon_{\bm{k}}}\nn&{}+\dfrac{\Sigma^z(\bm{k})}{4\varepsilon_{\bm{k}}}\bigg(\dfrac{\partial^2\Phi_{\bm{k}}^-}{\partial \varepsilon_{\bm{k}}^2}+\dfrac{\partial}{\partial \varepsilon_{\bm k}}\dfrac{\Phi_{\bm{k}}^-}{\varepsilon_{\bm{k}}}+\Delta^2\dfrac{\partial}{\partial \varepsilon_{\bm{k}}}\dfrac{1}{\varepsilon_{\bm{k}}}\dfrac{\partial}{\partial \varepsilon_{\bm k}}\dfrac{\Phi_{\bm k}^-}{\varepsilon_{\bm{k}}}\bigg)\nn&{}+\dfrac{\Delta\big(\bm{\Sigma^\parallel(\bm{k})\cdot\bm{k}}\big)}{4\varepsilon_{\bm{k}}}\dfrac{\partial}{\partial \varepsilon_{\bm{k}}}\dfrac{1}{\varepsilon_{\bm{k}}}\dfrac{\partial}{\partial \varepsilon_{\bm{k}}}\dfrac{\Phi^-_{\bm{k}}}{\varepsilon_{\bm{k}}}\bigg).
\end{align}

\subsection{\label{subsec:appendix_vertex}Diagrams in Fig.~\ref{fig:1}~(e)--(f)}

We now turn to diagrams depicted Fig.~\ref{fig:1}~(e)--(f) which arise because the microscopic energy-current and hence the heat-current operator contains an explicit interaction-dependent contribution, see second line of Eq. \eqref{eq:energy_current}. Unlike the self-energy, which dresses up the fermionic propagators, this term modifies the heat-current vertex itself and therefore constitutes a distinct first-order correction to the Kubo kernel. Such vertex corrections are absent in pure charge transport and therefore do not appear in the particle magnetisation, which involves only particle density and particle current density operators. They are instead essential for thermoelectric responses, where the energy current enters explicitly.
To $K_{xy}^{12}$, diagrams in Fig.~\ref{fig:1}~(e)--(f) contribute
\begin{align}
K_{xy}^{12(\mathrm{VC})}={}&\vf^2\Delta\int_{\bm{k}}\dfrac{\Phi^-_{\bm{k}}}{2\varepsilon_{\bm{k}}^3}\bigg(-\Sigma_0(\bm{k})\nn
&+\int_{\bm{k}'}V_{\bm{k}-\bm{k}'}\dfrac{\vf^2(\bm{k}-\bm{k}')\cdot\bm{k}'}{4\varepsilon_{\bm{k}'}}\dfrac{\partial \Phi^+_{\bm{k}'}}{\partial \varepsilon_{\bm{k}'}}\bigg).
\label{eq:appendix_K12_VC}
\end{align}

\section{\label{sec:appendix_combined_kubo}Summary of first-order contributions}

We can now collect the first-order interaction corrections to the thermoelectric Kubo kernel obtained from the exchange, self-energy, and vertex diagrams discussed in the preceding sections. Due to the Ward identity for the particle current, certain terms between the exchange contribution and the self-energy contribution are expected to cancel. That is why it is natural to sum these two first. For $K_{xy}^{12}$ the result is
\begin{align}
    K_\mathrm{I}={}&K_{xy}^{12(\mathrm{EX})}+K_{xy}^{12(\mathrm{SE})}\nn
    ={}&\dfrac{\mu\vf^2}{2}\int_{\bm{k}}\bigg(\dfrac{\Sigma^z(\bm{k})}{\varepsilon_{\bm{k}}^2}\dfrac{\partial \Phi^-_{\bm{k}}}{\partial \varepsilon_{\bm{k}}}-\dfrac{\Sigma^0(\bm{k})\Delta}{\varepsilon_{\bm{k}}^3}\dfrac{\partial \Phi_{\bm{k}}^+}{\partial \varepsilon_{\bm{k}}}\bigg)\nn
    &+\dfrac{\mu\Delta\vf^2}{4}\int_{\bm{k}\bm{k}'}V_{\bm{k}-\bm{k}'}\dfrac{\Phi^-_{\bm{k}}}{\varepsilon_{\bm{k}}^3}\dfrac{\partial \Phi^-_{\bm{k}'}}{\partial \varepsilon_{\bm{k}'}}\dfrac{(\bm{k}-\bm{k}')\cdot\bm{k}'}{\varepsilon_{\bm{k}'}^2}.
\end{align}
The rest of the diagrams we will put into $K_{\mathrm{II}}$ defined by
\begin{equation}
    K_{\mathrm{II}}=K_{xy}^{12(\mathrm{VC})},
\end{equation}
where $K_{xy}^{12(\mathrm{VC})}$ is given in Eq.~\eqref{eq:appendix_K12_VC}, so that $K=K_{\mathrm{I}}+K_{\mathrm{II}}$. The sum of the exchange and self-energy diagram contributions to $\partial M_N^z/\partial\mu$ undergoes massive cancellations and can be combined into
\begin{align}
    \dfrac{\partial (M_z^N)^{(1)}}{\partial \mu}=\vf^2\dfrac{\partial}{\partial\mu}\int_{\bm{k}}\bigg(\dfrac{\Sigma^z(\bm{k})}{2\varepsilon_{\bm{k}}}\dfrac{\partial \Phi_{\bm{k}}^+}{\partial \varepsilon_{\bm{k}}}-\dfrac{\Sigma_0(\bm{k})\Delta}{2\varepsilon_{\bm{k}}}\dfrac{\partial }{\partial \varepsilon_{\bm{k}}}\dfrac{\Phi_{\bm{k}}^-}{\varepsilon_{\bm{k}}}\bigg).\nn
\end{align}

\bibliographystyle{apsrev4-2}
\bibliography{References}
\end{document}